\documentclass[aps,pre,twocolumn,superscriptaddress,nofootinbib]{revtex4-2}
\usepackage{amsmath,amssymb}
\usepackage{graphicx}
\usepackage{siunitx}
\usepackage{booktabs}
\usepackage{xcolor}

\begin{document}

\title{Low-frequency output fluctuations in an open exclusion process with particle pausing}

\author{Quentin Thommen}
\affiliation{CRCLille--Cancer Research Center of Lille, Universit\'e de Lille, UMR 9020 CNRS, Inserm U1366, CHU de Lille, France}

\email{quentin.thommen@univ-lille.fr}

\date{\today}

\begin{abstract}
Slow internal states reshape both the mean throughput and the temporal organization of a driven lattice gas. Exit-counting statistics reveal this effect in a finite open totally asymmetric simple exclusion process whose particles reversibly switch between active and paused states. Increasing pausing lowers the mean current smoothly, whereas the long-window Fano factor is strongly nonmonotonic. At the reference boundary rates, the maximum remains near a measured mean paused population \(N_p=L\rho_{\rm paused}\simeq1.5\)--\(2\) across lattice lengths \(L=50\)--\(500\), while the corresponding pausing rate scales as \(k_p^{\rm max}\propto L^{-1}\). A minimal constant-birth, linear-death approximation translates an order-one collective crossover into this finite-size displacement and gives \(N_p^\star\simeq1.50\) in the independent-pause, strong-blocking limit. The simulations delimit this approximation: the pause number is overdispersed, and at fixed \(N_p\), slower unpausing increases both the correlation time and the noise amplitude. Residence-time and structural analyses further separate the relevant slow variables. The pause-free versus pause-containing residence-time scale tracks the fitted output-correlation time, whereas the noise amplitude follows fluctuations, rather than the mean size, of the largest particle cluster. Low-frequency output noise therefore identifies an intermittent finite-size regime shaped jointly by slow-defect kinetics and traffic-jam reorganization.
\end{abstract}

\maketitle

\section{Introduction}

Driven transport converts local stochastic events into collective fluctuations when particles interact through exclusion. The totally asymmetric simple exclusion process (TASEP), introduced to describe ribosome traffic \cite{MacDonald1968}, provides the canonical representation of this mechanism: particles move unidirectionally along a one-dimensional track, cannot overtake one another, and exchange with reservoirs at the boundaries \cite{Derrida1993,BlytheEvans2007,Chowdhury2013,Schadschneider2011}. A local interruption can therefore propagate upstream, reorganize a particle queue, and modify transport over the full system.

Transcription and translation add a slow internal state to this exclusion dynamics. RNA polymerases and ribosomes alternate between productive motion and pauses that range from short elemental events to rare arrested or backtracked states \cite{Davenport2000,Neuman2003,Landick2009}. A persistent pause acts as a dynamical roadblock under dense traffic: trailing particles accumulate, a finite jam develops, and the pause reduces throughput beyond the direct immobilization of the paused particle. Models of strongly transcribed genes have consequently linked pausing to broadened interparticle spacings, shocks, traffic jams, and reduced elongation currents \cite{KlumppHwa2008,Klumpp2011,SahooKlumpp2013}. The same collective dynamics can also reorganize the timing of production, generating convoys or clustered completion events and modifying downstream expression noise even without promoter switching \cite{Dobrzynski2009,Ribeiro2010,KimJacobs2018,Mines2022}.

The pausing TASEP introduced by Wang \textit{et al.} isolates this mechanism by allowing each particle to switch reversibly between active and paused states \cite{Wang2014}. Their current--density relation accounts for both paused particles and active particles blocked behind them, and thereby captures the reduction of the stationary current over a broad range of pause durations. Keisers \textit{et al.} subsequently developed the finite open-system theory of the same model \cite{Keisers2026PRE}. Their analysis derived the boundary-induced phase structure, demonstrated the equivalence between particle pausing and TASEPs with dynamical defects \cite{Turci2013}, and identified a strong finite-size regime when pausing and unpausing are slow relative to hopping. The mean number of paused particles then becomes the relevant finite-system variable. When this number is of order unity, pause-free intervals alternate with episodes in which one or a few pauses nucleate a large cluster. A single-cluster approximation describes how this alternation changes the mean current. The same physical picture has also been applied to transcript-length-dependent inhibition by ribosome-targeting antibiotics on degrading messenger RNAs \cite{Keisers2026PRXL}.

The finite-size regime is therefore established at the level of stationary transport, but its temporal output statistics remain unresolved. The stationary current fixes the average number of completed events per unit time; it does not distinguish a nearly regular output from productive episodes separated by long congested intervals. Integrated-current cumulants and large-deviation functions characterize such temporal fluctuations in exclusion processes \cite{DerridaLebowitz1998,LazarescuMallick2011,Gorissen2012,deGierEssler2011}, but the standard open ASEP does not contain the slow reversible particle state that generates the finite-size crossover considered here. Biological studies address related questions through different observables. Ali \textit{et al.}, for example, showed that the cell-to-cell Fano factor of the number of polymerases present on a gene becomes nonmonotonic when initiation is varied, with maximal variability when pause-induced bunch formation is sporadic \cite{Ali2020}. The resulting question is whether the finite-size pausing regime identified from the mean current possesses a distinct signature in the stationary statistics of completed exits.

The present work characterizes this signature through the number of particles leaving the lattice during an observation window of duration $T_w$. The corresponding Fano factor resolves the crossover from local counting fluctuations to slow output modulation. The directly measured mean number of paused particles,
\begin{equation}
    N_p=L\rho_{\rm paused},
\end{equation}
provides the collective control variable. Unlike the imposed rate $k_p$, $N_p$ measures the number of slow defects actually realized in the interacting open system and permits comparison between lattice lengths.

The output statistics reveal a fluctuation regime that remains largely hidden in the stationary current. At the reference boundary rates, the current decreases continuously as the paused population grows, whereas the low-frequency Fano factor reaches a pronounced maximum near $N_p\simeq1.5$--$2$ for all studied lengths. A minimal constant-birth, linear-death description rationalizes an order-one optimum and maps that collective crossover onto the finite-size scaling $k_p^{\rm max}\propto L^{-1}$. The simulations also define the limits of this approximation: the instantaneous pause number is not Poisson distributed, and $N_p$ organizes the position of the crossover at fixed boundary conditions but does not determine its amplitude. At comparable $N_p$, slower unpausing increases both the correlation time and the low-frequency plateau. Structural and residence-time measurements finally separate the underlying slow variables. The residence-time scale associated with the appearance and disappearance of paused particles tracks the dominant decorrelation time, whereas fluctuations of the largest jam provide the closest structural signature of the noise amplitude. Low-frequency output fluctuations therefore emerge from the coupled dynamics of slow defects and finite traffic jams rather than from a single exact telegraph state.

\section{Model, observables, and analytical framework}

\subsection{Open pausing TASEP}

\begin{figure}[t]
\includegraphics[width=\columnwidth]{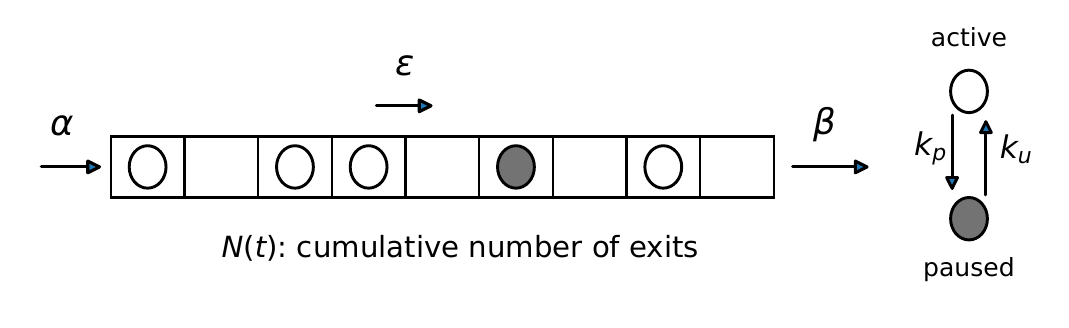}
\caption{Open pausing TASEP. Active particles hop with rate $\epsilon$ subject to exclusion, enter and leave the paused state with rates $k_p$ and $k_u$, enter with rate $\alpha$, and exit with rate $\beta$. $N(t)$ denotes the cumulative number of exit events.}
\label{fig:model}
\end{figure}

The model consists of a one-dimensional lattice of $L$ sites, each either empty or occupied by a single particle in an active state $A$ or a paused state $P$. An active particle at site $i<L$ hops to site $i+1$ with rate $\epsilon$, provided the target site is empty. Particles switch independently between internal states,
\begin{equation}
A \xrightarrow{k_p} P,
\qquad
P \xrightarrow{k_u} A.
\label{eq:internal}
\end{equation}
A particle enters site 1 with rate $\alpha$ when that site is empty, and an active particle at site $L$ exits with rate $\beta$; a paused particle at the final site must first resume activity before terminating. The counted output consists of these physical exit events. Figure~\ref{fig:model} summarizes the model and notation.

The active--paused transition rates fix the stationary internal-state fraction
\begin{equation}
f_p=\frac{k_p}{k_p+k_u}.
\end{equation}
Because these transitions act independently of position and exclusion, the same ratio applies to the stationary fractions of active and paused particles within the occupied population. The boundaries and congestion nevertheless determine the total occupancy, and therefore the absolute paused population. The directly measured quantity
\begin{equation}
N_p=L\rho_{\rm paused}=\langle N_{\rm paused}\rangle
\label{eq:Np}
\end{equation}
combines the internal kinetics with this interacting finite-system occupancy.

\subsection{Output counting and structural observables}

Let $N(t)$ denote the cumulative number of particles that have left the lattice up to time $t$. For an observation window of duration $T_w$,
\begin{equation}
N_w(t;T_w)=N(t+T_w)-N(t).
\end{equation}
The stationary output current is
\begin{equation}
J=\lim_{T_w\to\infty}\frac{\langle N_w\rangle}{T_w},
\end{equation}
and the window-dependent Fano factor is
\begin{equation}
F(T_w)=\frac{{\rm Var}(N_w)}{\langle N_w\rangle}.
\label{eq:fano}
\end{equation}
A Poisson counting process has $F=1$; the dependence on $T_w$ separates short-time exit statistics from slow modulation of the current.

The observed crossover is parametrized by
\begin{equation}
F(T_w)=F_0+(F_\infty-F_0)
\left[1-\frac{\tau_c}{T_w}\left(1-e^{-T_w/\tau_c}\right)\right].
\label{eq:telegraph}
\end{equation}
This expression is the integrated-count signature of an exponentially correlated current contribution; it does not imply an exact microscopic two-state process.

Output fluctuations connect to spatial organization through the sampled number of paused particles, the largest contiguous particle cluster $C_{\max}$, its position, the downstream-most paused particle, and the cluster attached to that pause. Residence-time diagnostics rely on thresholded congestion states defined from $C_{\max}/L$; the threshold-free quantities $\langle C_{\max}\rangle$ and ${\rm Var}(C_{\max})$ remain the primary structural observables.

\subsection{Minimal theory for the location of the noise maximum}

The simulations below show that \(N_p\), rather than the bare rate \(k_p\), organizes the maximum. Two approximations offer a minimal explanation of this observation.

In the slow-pausing regime \(k_p\ll k_u\ll\epsilon\),
\begin{equation}
f_p=\frac{k_p}{k_p+k_u}\simeq\frac{k_p}{k_u}.
\end{equation}
The relation between this single-particle fraction and the measured paused population follows directly from stationary pause balance. If \(N_A\) denotes the mean number of active particles on the lattice, stationarity of the number of paused particles requires
\begin{equation}
k_p N_A=k_uN_p.
\end{equation}
Since the mean total occupancy is \(L\rho=N_A+N_p\), this balance gives
\begin{equation}
N_p=L\rho f_p\simeq L\rho\frac{k_p}{k_u}.
\label{eq:Np_scaling}
\end{equation}
The exact stationary relation can also be inverted,
\begin{equation}
k_p=\frac{k_uN_p}{L\rho-N_p}.
\label{eq:kp_exact}
\end{equation}
Thus the conversion from the microscopic rate \(k_p\) to the collective variable \(N_p\) contains an explicit system-size dependence. If the total density changes only weakly near the noise maximum and the crossover occurs at a length-independent, order-one value \(N_p^\star\), then
\begin{equation}
k_p^{\rm max}
=\frac{k_uN_p^\star}{L\rho-N_p^\star}
\simeq\frac{k_uN_p^\star}{\rho L},
\qquad
k_p^{\rm max}\propto L^{-1},
\label{eq:kp_scaling}
\end{equation}
where the last form is the leading large-\(L\) limit \(N_p^\star\ll L\rho\). The scaling therefore does not assume that the bare pausing kinetics changes with system size: a longer lattice contains more particles exposed to pausing, so a smaller per-particle rate is sufficient to maintain the same order-one number of paused particles. Taking \(N_p^\star\simeq1.9\), \(k_u=10^{-3}\), and the measured density range \(\rho\simeq0.27\)--\(0.30\), the leading term in Eq.~\eqref{eq:kp_scaling} gives
\begin{equation}
Lk_p^{\rm max}\simeq6.3\times10^{-3}\text{--}7.0\times10^{-3}.
\end{equation}
The finite-\(L\) denominator in Eq.~\eqref{eq:kp_exact} produces the largest correction for the shortest lattice, while leaving the \(L^{-1}\) trend unchanged.

The second approximation addresses why an order-one paused population should maximize the slow noise. We replace the fluctuating aggregate pausing rate by its stationary mean \(\lambda=k_pN_A\) and approximate the instantaneous pause number by a constant-birth, linear-death process: pauses appear at rate \(\lambda\), while each paused particle resumes activity independently at rate \(k_u\). Stationary pause balance gives \(\lambda/k_u=N_p\), so this auxiliary process has a Poisson distribution with mean
\begin{equation}
n=\frac{\lambda}{k_u}=N_p.
\end{equation}
Its pause-free probability is therefore \(P_0=e^{-n}\). To retain only the slowest output modulation, the current is further reduced to
\begin{equation}
J(t)=J_p+(J_0-J_p)I_0(t),
\end{equation}
where \(I_0(t)\) indicates the absence of paused particles, and \(J_0\) and \(J_p\) are the effective currents in pause-free and pause-containing configurations.

For this immigration--death process,
\begin{equation}
{\rm Cov}[I_0(0),I_0(t)]
=e^{-2n}\left[\exp\!\left(ne^{-k_ut}\right)-1\right],
\end{equation}
and integrating gives
\begin{equation}
\int_0^\infty {\rm Cov}[I_0(0),I_0(t)]\,dt
=
\frac{e^{-2n}}{k_u}
\left[\operatorname{Ei}(n)-\gamma-\ln n\right].
\end{equation}
The corresponding slow contribution to the long-window Fano factor is
\begin{equation}
F_{\rm slow}(n)=
\frac{2(J_0-J_p)^2}{\overline{J}k_u}
e^{-2n}
\left[\operatorname{Ei}(n)-\gamma-\ln n\right],
\label{eq:Fslow}
\end{equation}
with
\begin{equation}
\overline{J}=J_p+(J_0-J_p)e^{-n}.
\end{equation}
In the strong-blocking limit \(J_p\ll J_0\), the \(n\)-dependent part reduces to
\begin{equation}
F_{\rm slow}(n)\propto
e^{-n}\left[\operatorname{Ei}(n)-\gamma-\ln n\right].
\end{equation}
Its nonzero maximum satisfies
\begin{equation}
\frac{e^n-1}{n}=\operatorname{Ei}(n)-\gamma-\ln n,
\end{equation}
giving
\begin{equation}
n^\star\simeq1.50.
\label{eq:nstar}
\end{equation}

This construction is not meant as a quantitative theory of the peak amplitude: it neglects the spatial position and finite lifetime of a nascent jam, correlations between pauses, boundary-dependent occupancy, and the continuous range of currents within pause-containing configurations. Its purpose is narrower: the approximation explains why maximal modulation occurs at an order-one mean number of slow defects, while stationarity maps this collective crossover onto an \(L^{-1}\) displacement of the microscopic pausing rate. The measured pause-number statistics below provide a direct test of its main approximation.

\subsection{Physical calibration and polymerase numbers}

Simulations express all rates in units of the hopping rate, setting $\epsilon=1$. Because the model treats particles as point-like, its mapping to molecular lengths is not unique; for order-of-magnitude comparison with bacterial transcription, one lattice site corresponds to roughly ten nucleotides -- a coarse graining of steric spacing, not a claim that a polymerase advances in ten-nucleotide elementary steps. A representative bacterial elongation velocity of $40~\mathrm{nt\,s^{-1}}$ then corresponds to
\begin{equation}
\epsilon\simeq4~\mathrm{s^{-1}},
\end{equation}
so one simulation time unit represents approximately $0.25~\mathrm{s}$. Under this coarse graining, $L=50$--$500$ represents transcription units of roughly $0.5$--$5~\mathrm{kb}$. For orientation, a $1~\mathrm{kb}$ transcription unit corresponds to $L\simeq100$, placing the simulated interval on the scale of individual bacterial genes and multi-kilobase transcription units. The values $\alpha=0.1$, $\beta=1$, and $k_u=10^{-3}$ correspond respectively to $0.4~\mathrm{s^{-1}}$, $4~\mathrm{s^{-1}}$, and $4\times10^{-3}~\mathrm{s^{-1}}$, the latter giving a mean pause duration of about $250~\mathrm{s}$.

The total number of particles on the lattice is
\begin{equation}
N_{\rm part}=L\rho.
\end{equation}
Density enters the finite-size mapping through the total occupancy and also sets the biological scale of the simulated particle numbers. Typical measured densities of order $0.15$--$0.35$ correspond to roughly $8$--$18$ particles for $L=50$, $15$--$35$ for $L=100$, $30$--$70$ for $L=200$, and $75$--$175$ for $L=500$. With the ten-nucleotide coarse graining, these values imply mean particle spacings of a few tens of nucleotides -- a dense-traffic regime representative of strongly transcribed units rather than a typical weakly expressed bacterial gene. This calibration establishes physically meaningful orders of magnitude, not a unique microscopic identification.

\subsection{Simulation protocol}

The dynamics is simulated exactly in continuous time using an
event-driven kinetic Monte Carlo algorithm. Unless otherwise
stated, $\epsilon=1$, $\alpha=0.1$, $\beta=1$, and
$k_u=10^{-3}$; the pausing rate $k_p$ and the lattice length
$L$ are varied. A dedicated boundary-rate control uses $L=100$ and $\alpha=0.03$ while keeping the other reference rates unchanged. Each parameter point of the main Fano-factor
scan is sampled through ten independent trajectories after a
rate-adapted relaxation period. Exit-count statistics,
structural observables, residence times, fitting procedures,
and uncertainty estimates are described in
Appendix~\ref{app:numerical_methods}.

\section{Results}

\subsection{Intermittent output and observation-time-dependent noise}

\begin{figure}[t]
\centering
\includegraphics[width=\columnwidth]{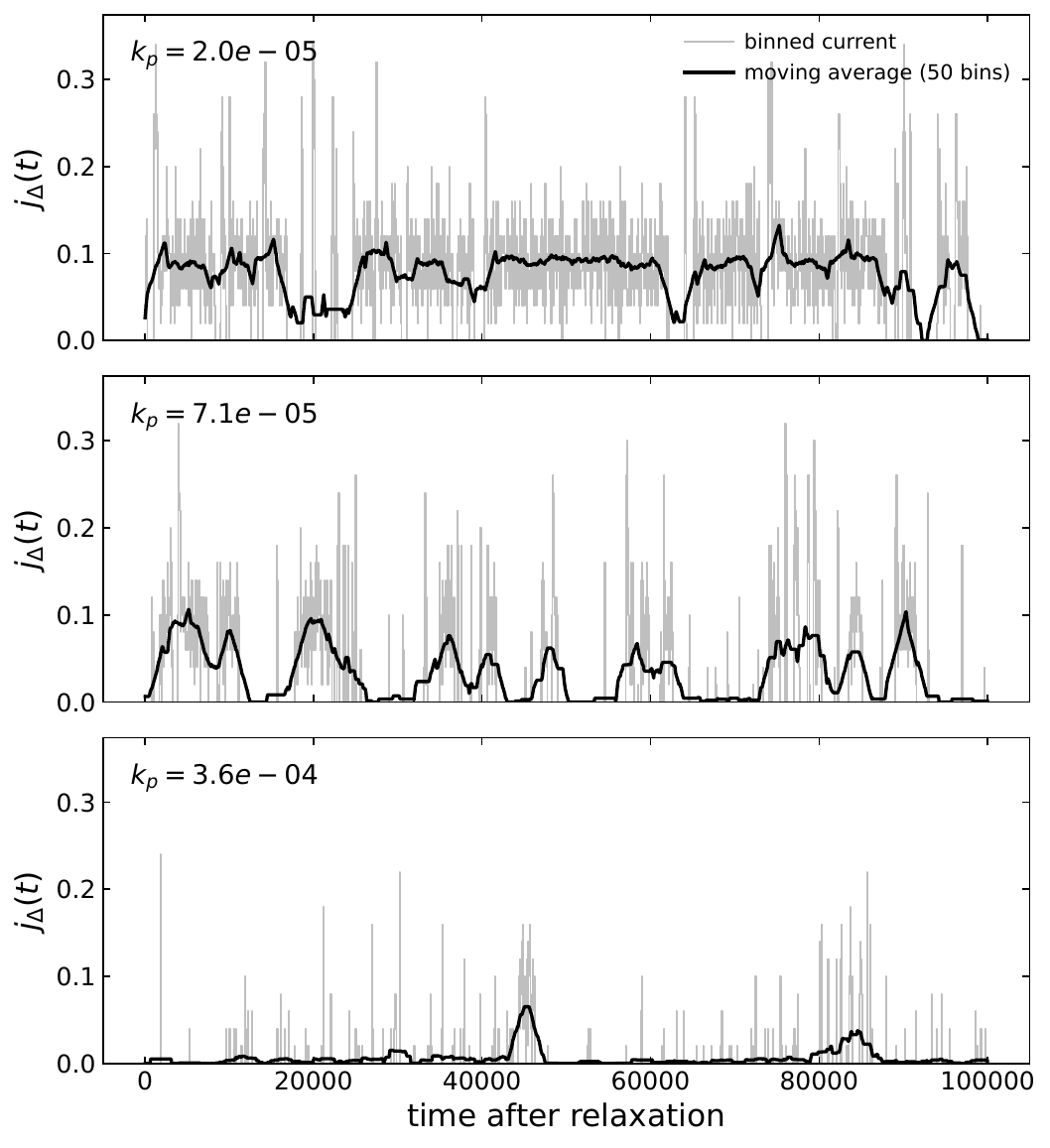}
\caption{Event-resolved output time series for \(L=100\), \(k_u=10^{-3}\), \(\alpha=0.1\), and \(\beta=\epsilon=1\), in three representative pausing regimes. The three panels use identical time and current scales. The black curve is a moving average included only as a visual guide.}
\label{fig:timeseries}
\end{figure}

Event-resolved trajectories reveal the qualitative change in the exit process. Figure~\ref{fig:timeseries} shows the binned current
\begin{equation}
j_\Delta(t)=\frac{N(t+\Delta)-N(t)}{\Delta}
\end{equation}
for three representative \(L=100\) simulations, with identical axes in all panels. For rare pauses ($k_p=2\times 10^{-5}$) the output stays comparatively regular, with infrequent interruptions. Near the noise maximum ($k_p=7.1\times 10^{-5}$), productive intervals alternate with extended periods of weak or vanishing output, and at larger paused populations ($k_p=3.6\times 10^{-4}$) the current stays suppressed for most of the trajectory. The intermediate regime shows the largest temporal contrast between configurations visited along a stationary trajectory, rather than the smallest current.

\begin{figure}[t]
\includegraphics[width=\columnwidth]{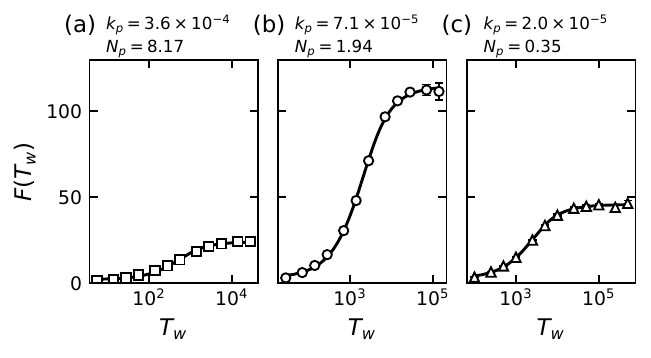}
\caption{Window-dependent Fano factor for representative \(L=100\) simulations: (a) frequent pauses, (b) the noise-maximum regime, and (c) rare pauses. Symbols show simulation results and lines are fits to Eq.~\eqref{eq:telegraph}. The intermediate paused population produces the largest low-frequency plateau.}
\label{fig:fano}
\end{figure}

Figure~\ref{fig:fano} quantifies these regimes by changing the observation time rather than the microscopic parameters. At short $T_w$, each window probes only local exit statistics and $F(T_w)$ remains close to a short-time counting baseline. Increasing $T_w$ allows a single count to integrate over the slow alternation between productive and inhibited intervals; the resulting extra covariance raises $F(T_w)$ until windows become long compared with the correlation time and the Fano factor approaches a plateau. The height of this plateau, $F_\infty$, measures the strength of the slow output modulation, while the crossover along the horizontal axis defines its effective duration $\tau_c$. The intermediate pausing regime therefore stands out in two complementary ways: it produces the largest long-time variance relative to the mean, and it does so through fluctuations that remain correlated over a finite slow timescale. Equation~\eqref{eq:telegraph} captures both features over the sampled range and provides a common parametrization for comparing $F_\infty$ and $\tau_c$ across parameters.

These data establish the distinction pursued below: increasing pausing continuously lowers the mean throughput, while relative low-frequency noise varies nonmonotonically with the pausing rate. A large \(F_\infty\) requires repeated sampling of configurations with substantially different output currents.

\subsection{Finite-size organization of the noise maximum}

\begin{figure}[t]
\includegraphics[width=\columnwidth]{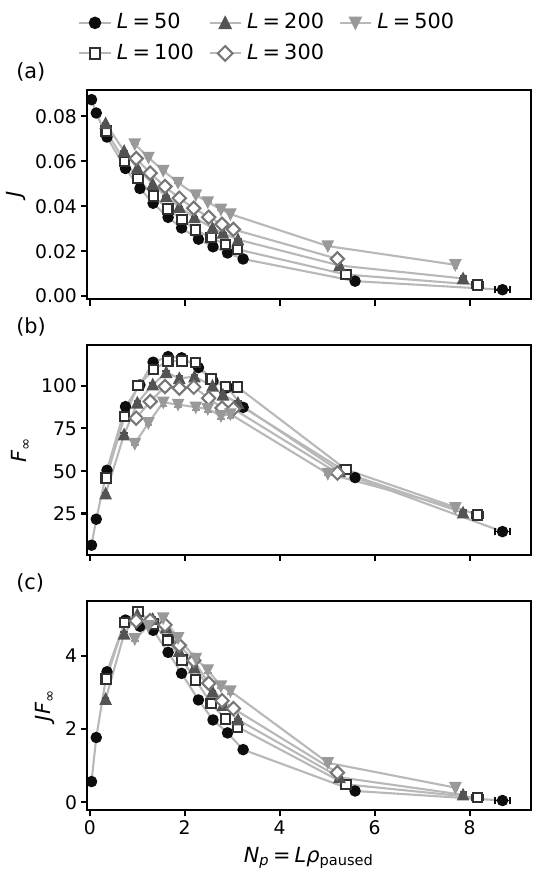}
\caption{Mean output and low-frequency fluctuations organized by the measured paused population. (a) Stationary current \(J\). (b) Fitted long-window Fano plateau \(F_\infty\), which measures fluctuations relative to the mean output and peaks near \(N_p\simeq1.5\)--\(2\). (c) Absolute long-window fluctuation rate \(J F_\infty=\lim_{T_w\to\infty}\mathrm{Var}[N_w]/T_w\). Its pronounced nonmonotonicity shows that the maximum in relative noise is not generated solely by division by a decreasing current. The maximum of \(J F_\infty\) occurs at a somewhat smaller paused population, of order unity, and its peak amplitude varies only weakly with lattice length over the simulated range.}
\label{fig:collapse}
\end{figure}

Figure~\ref{fig:collapse} separates the mean output, the relative low-frequency noise, and the corresponding absolute fluctuation rate at the reference initiation rate $\alpha=0.1$. The stationary current decreases throughout the scan, whereas \(F_\infty\) rises sharply, peaks near \(N_p\simeq1.5\)--\(2\), and then falls. The product \(J F_\infty\), equal to the long-window variance growth rate \(\lim_{T_w\to\infty}\mathrm{Var}[N_w]/T_w\), is itself strongly nonmonotonic and reaches a maximum at a smaller paused population, typically \(N_p=O(1)\). The maximum of \(F_\infty\) therefore does not arise solely from normalization by the decreasing mean current: absolute output fluctuations are also maximal in the intermittent finite-size regime.

Table~\ref{tab:maxima} summarizes the four lengths for which a complete local scan resolved the maximum. The peak position shifts only weakly, from \(N_p^{\rm max}=1.94\) at \(L=50\) and \(100\) to \(1.86\) at \(L=500\), even as \(k_p^{\rm max}\) changes by more than one decade; the product \(Lk_p^{\rm max}\) stays approximately constant, consistent with Eq.~\eqref{eq:kp_scaling}. The finite-size scans therefore give
\begin{equation}
k_p^{\rm max}\propto L^{-1},
\end{equation}
with a maximum located at an order-one paused population.

\begin{table}[t]
\caption{Location and magnitude of the low-frequency noise maximum.}
\label{tab:maxima}
\begin{ruledtabular}
\begin{tabular}{ccccc}
\(L\) & \(k_p^{\rm max}\) & \(N_p^{\rm max}\) & \(F_\infty^{\rm max}\) & \(k_u\tau_c\)\\
\hline
50  & \(1.50\times10^{-4}\) & 1.94 & 116 & 1.06\\
100 & \(7.10\times10^{-5}\) & 1.94 & 115 & 1.19\\
200 & \(3.40\times10^{-5}\) & 1.89 & 104 & 1.23\\
500 & \(1.35\times10^{-5}\) & 1.86 & 89  & 1.04\\
\end{tabular}
\end{ruledtabular}
\end{table}

At the maxima, \(k_u\tau_c=O(1)\), pointing to an unpausing-related decorrelation scale, though the peak amplitude decreases for the largest systems. \(N_p\) therefore stands as a scaling variable for the \emph{location} of the crossover, rather than a complete collapse variable for \(F_\infty\). A separate boundary-rate control first tests whether this organization persists when initiation is reduced.

A control scan at $L=100$ reduces the initiation rate from $\alpha=0.1$ to $0.03$ while keeping $\beta=\epsilon=1$ and $k_u=10^{-3}$. The relative-noise maximum remains pronounced but shifts to $k_p=1.6\times10^{-4}$ and $N_p=3.26$, where $F_\infty=47.25$, with neighboring values $F_\infty=40.41$ at $N_p=1.57$ and $42.15$ at $N_p=4.51$. The absolute fluctuation rate $JF_\infty$ is also nonmonotonic and reaches its sampled maximum at $N_p=1.57$. Reducing initiation therefore preserves the intermittent fluctuation regime and its order-one paused-population scale, while shifting its precise location and amplitude. Appendix~\ref{app:alpha_control} gives the control protocol and full scan. The following scan then separates pause abundance from pause lifetime by varying $k_u$ at comparable $N_p$.

\subsection{Unpausing sets the time scale and changes the noise amplitude}

The scans above, performed at a fixed unpausing rate $k_u=10^{-3}$, showed that $N_p$ organizes the position of the low-frequency noise maximum. Whether it also determines the full output statistics is a separate question. A dedicated scan varies $k_u$ at fixed lattice length $L=100$, adjusting $k_p$ to hold $N_p$ roughly constant, for five unpausing rates $k_u=5\times10^{-4}$ to $5\times10^{-3}$ and three target populations $N_p^{\rm target}=1.5$, $2$, and $3$.

\begin{figure}[t]
\includegraphics[width=\columnwidth]{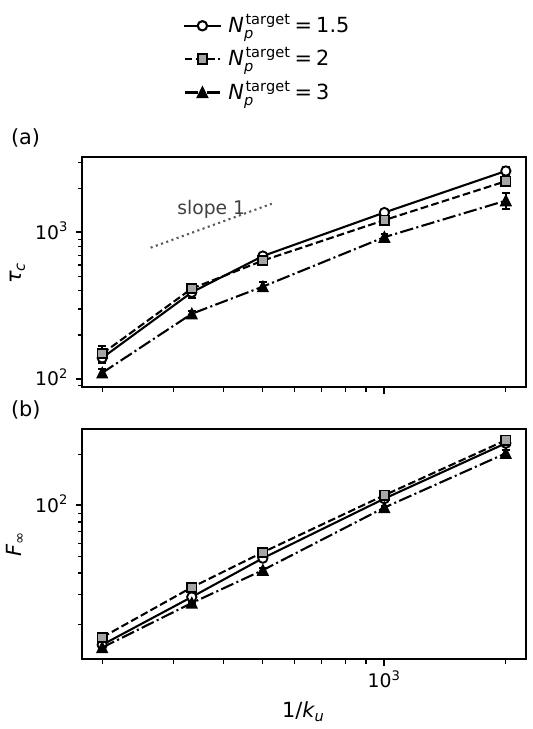}
\caption{Effect of the unpausing rate at $L=100$ for three
target paused populations. The pausing rate $k_p$ varies
with $k_u$ to maintain comparable measured
values of $N_p$. (a) Fitted output-correlation time as a
function of $1/k_u$. The dashed guide has unit slope.
(b) Fitted low-frequency plateau $F_\infty$ as a function of
$1/k_u$. Both the correlation time and the noise amplitude
increase when unpausing becomes slower, showing that $N_p$
alone leaves the output statistics underdetermined. Lines are
effective power-law fits over the simulated range.}
\label{fig:ku_scan}
\end{figure}

Figure~\ref{fig:ku_scan}(a) tests whether the output-correlation time follows the pause lifetime. A unit slope against $1/k_u$ corresponds to $\tau_c\propto k_u^{-1}$. The three target populations lie close to this guide over most of the scanned decade: effective fits give $\tau_c\propto k_u^{-\gamma_\tau}$ with $\gamma_\tau\simeq1.1$--$1.2$. Equivalently, $k_u\tau_c$ remains of order unity except at the largest $k_u$. This near-proportionality identifies unpausing as the dominant microscopic clock for the slow output modulation, while the residual departures from unit slope leave room for additional traffic-relaxation times.

Figure~\ref{fig:ku_scan}(b) then asks whether fixing $N_p$ also fixes the magnitude of that modulation. It does not. At $N_p\simeq2$, decreasing $k_u$ from $5\times10^{-3}$ to $5\times10^{-4}$ raises $F_\infty$ from about $17$ to $242$, although the measured paused population changes only from $1.85$ to $1.95$. The three target populations show the same systematic increase with $1/k_u$, summarized over this finite range by effective fits $F_\infty\propto k_u^{-\gamma_F}$ with $\gamma_F\simeq1.15$. Thus two systems containing essentially the same mean number of paused particles can have very different low-frequency output noise if those pauses persist for different durations.

These results refine the role of $N_p$: it robustly sets the position of the finite-size crossover at fixed $k_u$, but is not a complete state variable for the amplitude of the output noise. Slower unpausing lets pause-induced clusters persist and reorganize the traffic over longer times, raising both the correlation time and the contrast between productive and inhibited intervals. The low-frequency noise therefore depends on both the number of paused particles and the ratio between the pause lifetime and the characteristic times of cluster growth and relaxation.

\subsection{Microscopic organization of the intermittent regime}
\label{sec:microscopic_organization}

The preceding results establish where the low-frequency noise is maximal and which kinetic parameters control its position and timescale. They do not yet identify which microscopic configurations carry this excess variability. The structural analysis therefore addresses three successive questions. First, does the noise follow the mean extent of congestion or its temporal fluctuations? Second, can the crossover be reduced to the statistics of the number of paused particles? Third, which microscopic partition of the dynamics reproduces the fitted correlation time? This progression separates the spatial amplitude of the intermittency from the slow process that sets its duration.

\subsubsection{Fluctuations of the dominant jam}

\begin{figure}[t]
\includegraphics[width=\columnwidth]{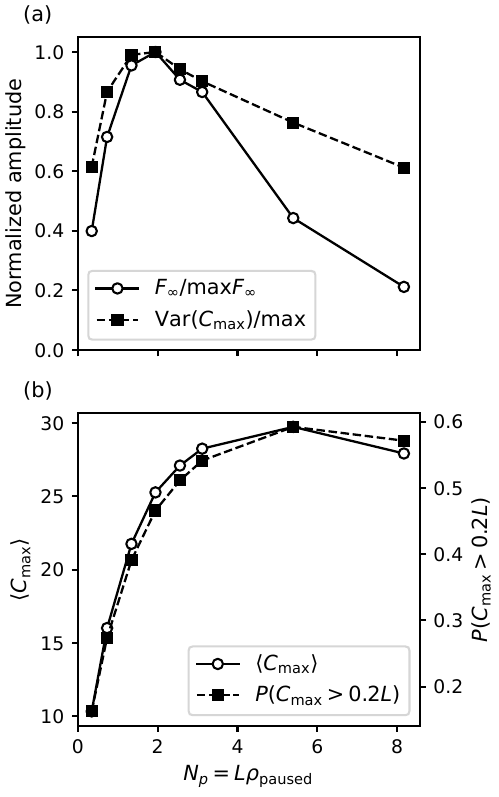}
\caption{Structural organization of the output-noise maximum for $L=100$. (a) Low-frequency Fano factor and variance of the largest cluster, each normalized by its maximum value. (b) Mean largest-cluster size and probability of satisfying $C_{\max}>0.2L$. The fluctuation amplitude is nonmonotonic, whereas the mean size and blocked-state probability remain large beyond the noise maximum.}
\label{fig:structure}
\end{figure}

The structural scan at $L=100$ first tests whether the output noise follows the mean congestion or the variability of the dominant jam. Figure~\ref{fig:structure}(a) compares $F_\infty$ with the variance of the largest cluster. Both quantities peak near $N_p\simeq1.5$--$2$. Across the eight sampled parameter sets, the descriptive Pearson correlation between $F_\infty$ and ${\rm Var}(C_{\max})$ is approximately $0.97$ ($n=8$). Fluctuations of the cluster attached to the downstream-most pause show a comparable association.

The mean largest-cluster size follows a different trend. It grows through the crossover and remains large after $F_\infty$ has begun to decrease; the probability of $C_{\max}>0.2L$ also stays high at larger $N_p$ [Fig.~\ref{fig:structure}(b)]. Persistent congestion therefore explains the suppression of the mean current, but not the nonmonotonic output noise. The noise is largest when the dominant cluster explores a broad range of sizes, consistent with repeated growth, shrinkage, and reorganization of the jam.

This distinction organizes the structural dynamics into three regimes. For $N_p\ll1$, pauses rarely persist long enough to generate large clusters, so the output remains comparatively regular. For $N_p=O(1)$, the system alternates between weakly and strongly congested configurations, and the variance of the largest cluster reaches its maximum. For $N_p\gg1$, congestion becomes persistent; the current stays low, but the system explores a narrower set of uniformly inhibited configurations and the relative output noise decreases.

The cluster analysis identifies the structural observable most closely associated with the noise amplitude. It does not yet explain why the crossover occurs at an order-one paused population. The instantaneous distribution of paused particles provides the next test of that finite-size interpretation.

\subsubsection{Pause-number statistics}
\label{sec:pause_number_statistics}

The analytical description assumes that the instantaneous number of paused particles, $M=N_{\mathrm{paused}}(t)$, follows a Poisson distribution with mean $N_p$. Under this independent-pause approximation,
\begin{equation}
P(M=0)=e^{-N_p},
\qquad
P(M=1)=N_p e^{-N_p},
\end{equation}
and
\begin{equation}
P(M\geq2)=1-e^{-N_p}(1+N_p).
\end{equation}

\begin{figure}[t]
    \centering
    \includegraphics[width=\columnwidth]{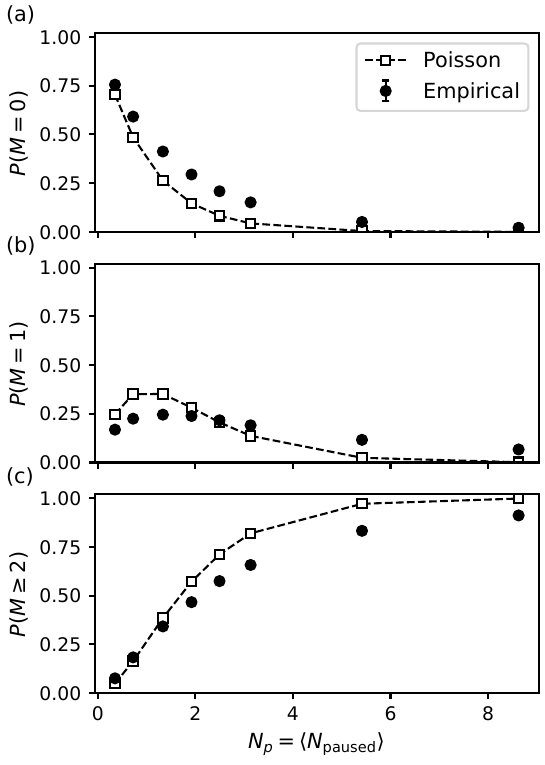}
    \caption{
    Instantaneous pause-number statistics for $L=100$,
    $\alpha=0.1$, $\beta=\epsilon=1$, and $k_u=10^{-3}$.
    Filled circles show the empirical fractions of time spent in
    configurations with (a) no paused particle, (b) exactly one
    paused particle, and (c) two or more paused particles.
    Open squares connected by dashed lines show the corresponding
    predictions of a Poisson distribution with the same measured
    mean $N_p=\langle N_{\mathrm{paused}}\rangle$:
    $e^{-N_p}$, $N_p e^{-N_p}$, and
    $1-e^{-N_p}(1+N_p)$, respectively.
    Error bars represent circular moving-block-bootstrap
    uncertainties computed from the stored structural
    trajectory at each parameter point and are generally
    smaller than the symbols.
    The Poisson approximation
    reproduces the qualitative crossover but underestimates the
    probability of pause-free configurations and overestimates the
    probability of configurations containing several pauses.
    Near the output-noise maximum, $N_p\simeq1.5$--$2$, all three
    classes retain substantial statistical weight.
    }
    \label{fig:pause_classes}
\end{figure}

Figure~\ref{fig:pause_classes} compares these expressions with the empirical fractions of time spent in the three pause-number sectors. The approximation reproduces the overall progression from predominantly pause-free configurations at small $N_p$ to configurations with several pauses at large $N_p$. The quantitative deviations are systematic, however: pause-free configurations occur more often than $e^{-N_p}$ predicts, whereas configurations with two or more pauses occur less often than their Poisson estimate. The pause number is also overdispersed, with $\operatorname{Var}(M)/\langle M\rangle>1$ throughout the sampled range.

At the output-noise maximum, $N_p\simeq1.92$, the measured probabilities are $P(M=0)\simeq0.295$, $P(M=1)\simeq0.239$, and $P(M\geq2)\simeq0.467$. None of the three sectors dominates. Pause-free, single-pause, and multiple-pause configurations all retain substantial weight, which places the maximum in a genuine crossover regime rather than in a sharply defined single-pause state.

A five-trajectory reproducibility control at the same regime gives $P(M=0)=0.291\pm0.004$, $P(M=1)=0.234\pm0.004$, and $P(M\geq2)=0.474\pm0.006$ (standard deviations across independent trajectories), confirming that the plotted structural trajectory is representative.

The deviations from Poisson statistics also delimit the analytical approximation. The mean paused population captures where the different sectors coexist, but it does not determine their full probability distribution. Occupancy fluctuations, altered particle residence times, and exclusion-induced congestion feed back on pause accumulation and produce correlations absent from the independent immigration--death process.

Across the eight sampled parameter sets, $P(M=1)$ is strongly associated with $F_\infty$, but ${\rm Var}(C_{\max})$ remains the closer structural correlate. The pause-number statistics therefore explain why the crossover occurs when the mean paused population is of order unity, whereas fluctuations of the dominant jam more directly describe how strongly the output is modulated.

The pause-number sectors clarify the location of the crossover, but they do not yet identify the timescale measured by $\tau_c$. The final step therefore compares the fitted output-correlation time with residence times constructed from two candidate coarse-grained descriptions: congestion defined by the largest cluster, and pause activity defined by the presence of at least one paused particle.

\subsubsection{Residence times and effective telegraph variables}
\label{sec:telegraph_residence}

The empirical form used to fit $F(T_w)$ suggests an effective two-state modulation of the output. This interpretation does not specify which microscopic variable defines the two states. We therefore compare the fitted correlation time $\tau_c$ with residence-time scales derived from the structural trajectories.

\begin{figure}[t]
    \centering
    \includegraphics[width=0.8\columnwidth]{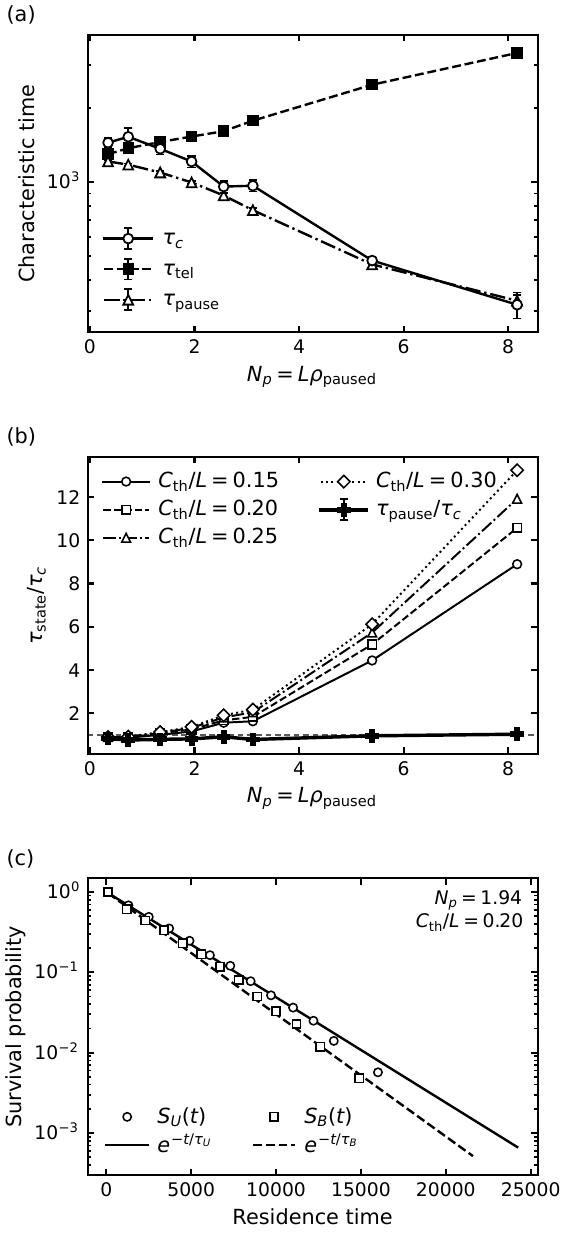}
    \caption{Comparison between output-correlation times and
      residence-time estimates for $L=100$, $\alpha=0.1$,
      $\beta=\epsilon=1$, and $k_u=10^{-3}$.  (a) Correlation time
      $\tau_c$ obtained from fits of $F(T_w)$, telegraph time
      $\tau_{\mathrm{tel}}$ obtained from weakly and strongly
      congested states defined by $C_{\max}/L=0.20$, and
      residence-time scale $\tau_{\mathrm{pause}}$ obtained from the
      states $N_{\mathrm{paused}}=0$ and $N_{\mathrm{paused}}\geq1$.
      (b) Ratio of the structural residence-time estimates to
      $\tau_c$. The cluster-threshold estimate becomes increasingly
      larger than $\tau_c$ as $N_p$ grows, independently of the
      precise threshold, whereas $\tau_{\mathrm{pause}}/\tau_c$
      remains of order unity and varies only weakly across the
      crossover.  Residence-time error bars are moving-block-bootstrap
      standard deviations computed from the stored structural
      trajectory at each parameter point.  (c) Survival probabilities
      of the weakly and strongly congested residence times near the
      noise maximum, $N_p=1.94$, for $C_{\mathrm{th}}/L=0.20$.  Lines
      show exponential distributions with the same mean residence
      times. The approximate but imperfect agreement supports an
      effective, rather than exact, telegraph description.  }
    \label{fig:telegraph_residence}
\end{figure}

A first partition separates weakly and strongly congested configurations through a threshold on the largest cluster,
\begin{equation}
U:\ C_{\max}\leq C_{\mathrm{th}},
\qquad
B:\ C_{\max}>C_{\mathrm{th}}.
\end{equation}
For an exact two-state Markov process, the corresponding correlation time would be
\begin{equation}
\tau_{\mathrm{tel}}
=
\frac{\tau_U\tau_B}{\tau_U+\tau_B},
\end{equation}
where $\tau_U$ and $\tau_B$ are the mean residence times in the two states. Figure~\ref{fig:telegraph_residence}(a) shows that this cluster-threshold timescale does not track $\tau_c$. The two timescales are comparable at small $N_p$, but $\tau_{\mathrm{tel}}$ increases with the paused population while $\tau_c$ decreases. Moderate changes in $C_{\mathrm{th}}/L$ preserve this disagreement [Fig.~\ref{fig:telegraph_residence}(b)]. A fixed threshold on $C_{\max}$ therefore does not define the slow state variable underlying the output decorrelation.

A second partition uses only the presence of paused particles,
\begin{equation}
U_0:\ N_{\mathrm{paused}}=0,
\qquad
P:\ N_{\mathrm{paused}}\geq1.
\end{equation}
The corresponding residence-time scale,
\begin{equation}
\tau_{\mathrm{pause}}
=
\frac{\tau_0\tau_P}{\tau_0+\tau_P},
\end{equation}
remains close to the independently fitted $\tau_c$ over the scanned range. Within the temporal resolution of the structural sampling, the appearance and disappearance of paused particles therefore capture the dominant output-decorrelation time more closely than crossings of a fixed congestion threshold. Five-trajectory controls at $N_p\simeq0.73$, $1.95$, and $3.12$ preserve this distinction: the replicate-wise ratio $\tau_{\rm pause}/\tau_c$ remains approximately constant near $0.8$, whereas $\tau_{\rm tel}/\tau_c$ for $C_{\rm th}/L=0.20$ increases from about $1$ to $1.8$ across the same conditions (Appendix~\ref{app:residence_times}).

Near the noise maximum, the survival probabilities of the weakly and strongly congested states are approximately exponential over a substantial range of residence times [Fig.~\ref{fig:telegraph_residence}(c)]. The remaining deviations preclude an exact Markov reduction, but they support the use of the telegraph form as an effective parametrization of the slow output modulation.

The three structural analyses now separate distinct roles for the microscopic variables. Pause-number dynamics identifies the finite-size crossover and provides the residence-time scale that most closely tracks the fitted output-correlation time. Largest-cluster fluctuations, by contrast, provide the closest structural correlate of the noise amplitude. These observables therefore describe complementary levels of the same intermittent dynamics: pauses initiate slow episodes, while the ensuing jam reorganization is associated with how strongly those episodes perturb the output.

\clearpage
\section{Discussion}

The present results reveal a finite-size fluctuation regime that the stationary current alone does not resolve. Increasing the paused population progressively suppresses the mean output, but the low-frequency Fano factor reaches its maximum only in an intermediate regime. Persistent congestion therefore does not imply maximal relative noise. A large Fano factor instead requires the traffic to alternate repeatedly between configurations with sufficiently different output rates.

The measured paused population provides the first level of this organization. For $N_p\ll1$, the lattice is usually pause-free, and pause-induced jams remain too rare to dominate long-time counting statistics. When $N_p=O(1)$, pause-free, single-pause, and multiple-pause configurations all retain appreciable statistical weight. The traffic can then explore both productive and congested configurations, and the dominant cluster spans its broadest range of sizes. For $N_p\gg1$, congestion persists over most of the trajectory. The mean current remains small, but the system samples a narrower family of inhibited configurations and the relative output noise decreases. The exit traces, the pause-sector probabilities, and the maximum of $\operatorname{Var}(C_{\max})$ converge on this three-regime interpretation.

This fluctuation picture extends the mean-current descriptions of Wang \textit{et al.} and Keisers \textit{et al.} \cite{Wang2014,Keisers2026PRE}. Those studies established the reversible pausing mechanism, the formation of a dominant cluster, and the importance of an order-one paused population for finite systems. The present results show that the same regime produces a sharper signature in the exit process: at the reference boundary rates, the low-frequency noise is nonmonotonic, its maximum remains near $N_p\simeq1.5$--$2$ over the studied lengths, and the pausing rate required to reach this regime scales as $L^{-1}$. The stationary pause balance makes the last relation explicit: once an absolute number of slow defects controls the crossover, increasing $L$ mainly reduces the microscopic pausing rate required to maintain that population.

The relation to Ali \textit{et al.} \cite{Ali2020} is complementary. Their analysis concerns the cell-to-cell Fano factor of the instantaneous number of polymerases on a gene while initiation is varied. The present analysis instead concerns the temporal counting statistics of completed exits while the pause and unpause rates are varied at fixed boundaries. Both observables detect intermittent bunching, but they answer different biological questions. Polymerase-number fluctuations report how elongation reorganizes occupancy on the template. Exit fluctuations report how completed RNA molecules are supplied over time.
Exit-counting statistics directly characterize the temporal delivery of completed transcripts, whereas polymerase-number statistics characterize occupancy of the template. These observables therefore probe complementary aspects of the same elongation dynamics and need not share the same fluctuation structure. For processes sensitive to transcript completion times, the exit statistics provide the relevant temporal input.

The minimal constant-birth, linear-death description captures the leading finite-size mechanism without reproducing the full interacting dynamics. It places the independent-pause, strong-blocking maximum at $N_p^\star\simeq1.50$ and, through stationary pause balance, maps an order-one crossover onto $k_p^{\rm max}\propto L^{-1}$ at fixed boundary conditions. The empirical pause statistics nevertheless depart systematically from the underlying Poisson assumption: the pause number is overdispersed, pause-free configurations occur more frequently than $e^{-N_p}$, and configurations containing several pauses occur less frequently than the independent-pause estimate. The numerical maximum near $N_p\simeq1.9$ should therefore be interpreted as an emergent value for the reference boundary conditions and hierarchy of time scales, rather than as a universal constant. This interpretation is supported directly by the $\alpha=0.03$ control: the pronounced maximum persists but shifts to $N_p\simeq3$ while remaining at an order-one paused population. Finite jam-formation times, pause positions, boundary-driven density changes, occupancy feedback, and exclusion-induced correlations provide natural corrections to the minimal theory.

The time-resolved analyses further separate the slow variables involved in the output modulation. Equation~\eqref{eq:telegraph} provides a compact empirical description of $F(T_w)$, and the residence-time scale obtained by distinguishing pause-free from pause-containing configurations remains approximately proportional to the fitted $\tau_c$. Five-trajectory controls keep $\tau_{\rm pause}/\tau_c$ near $0.8$ across three conditions spanning the crossover, whereas the cluster-threshold ratio increases systematically with $N_p$. This behavior supports the view that the appearance and disappearance of paused particles track the dominant decorrelation time. A binary variable constructed by thresholding $C_{\max}$ does not reproduce the same time scale once $N_p$ increases. A jam can grow, shrink, or reorganize while remaining on one side of a fixed threshold, so the threshold-crossing time need not coincide with the correlation time of the exit current.

The largest cluster nevertheless remains central to the noise amplitude. Its mean size and the probability of exceeding a threshold stay large beyond the maximum of $F_\infty$, whereas $\operatorname{Var}(C_{\max})$ follows the nonmonotonic noise much more closely. These observations suggest a two-layered effective description. Pause-number dynamics provides the natural clock for output decorrelation, while the range of structural reorganizations of the dominant jam is associated with the strength of current modulation. The data support this separation, although they do not yet provide a closed analytical expression for $F_\infty$.

The scan in $k_u$ reinforces this limitation of $N_p$ as a complete control variable. At comparable measured $N_p$, slower unpausing increases both $\tau_c$ and $F_\infty$. The increase in $\tau_c$ follows naturally from the longer pause lifetime. The increase in amplitude indicates that longer pauses also modify the traffic state reached during an inhibited episode, presumably by allowing the dominant jam to grow and reorganize over a larger range. A quantitative theory of the plateau amplitude will therefore require an additional dimensionless ratio comparing the pause lifetime with the characteristic time for jam formation and relaxation. The residual size dependence of the maximum likely reflects the same unresolved dynamical competition.

The biological interpretation should consequently focus on temporal delivery rather than on average occupancy alone. A given mean transcription rate can arise from a nearly regular sequence of completions or from productive episodes separated by extended silent intervals. These output patterns expose downstream processes to different temporal inputs even when their long-time means coincide. RNA degradation can erase or preserve these fluctuations depending on transcript lifetime, translation can convert them into protein-number variability, and regulatory networks can amplify them when their intrinsic response times overlap the output-correlation time. The present model therefore identifies a regime in which elongation traffic can shape gene-expression dynamics without promoter switching or sequence-specific pausing.

The effective calibration places this regime within plausible, but restricted, transcriptional scales. The simulated lengths correspond to approximately $0.5$--$5~\mathrm{kb}$, and the occupancies represent strongly transcribed units. The mapped mean pause time for $k_u=10^{-3}$ is about $250~\mathrm{s}$, which is more consistent with a rare arrested, backtracked, or obstacle-induced state than with a typical elementary pause. The model therefore applies most directly to dense traffic exposed to rare and persistent interruptions.

Several simplifications delimit the scope of the conclusions. Particles are point-like, hopping and pausing are spatially homogeneous, and the dynamics excludes promoter switching, sequence-dependent pause sites, backtracking, premature termination, transcript degradation, and template heterogeneity. These ingredients may shift the peak position, alter its amplitude, or add independent low-frequency components. Structural measurements were performed at $L=100$, so their relation to the residual size dependence of the amplitude remains to be established. The finite-size scans use an initiation-limited, non-exit-limited reference regime, and the $\alpha=0.03$ control shows explicitly that the numerical value of $N_p^\star$ shifts when the boundary-driven density changes. The peak position therefore carries no claim of universality across the full open-system phase diagram. Within these limits, the homogeneous model isolates a fluctuation maximum generated by collective elongation traffic alone.

\section{Conclusion}

Reversible pausing in a finite open exclusion process separates mean transport from temporal delivery. The mean current decreases progressively with the paused population, whereas the low-frequency output noise reaches a maximum when the mean number of paused particles is of order unity. This fluctuation maximum persists when initiation is reduced more than threefold, although its precise location shifts with the boundary-driven density. A minimal analytical description rationalizes the order-one crossover and its associated $L^{-1}$ displacement of the microscopic pausing rate at fixed boundary conditions. The simulations then resolve the collective corrections: the residence-time scale of pause-number dynamics tracks the dominant correlation time, while fluctuations of the largest jam provide the closest structural signature of the noise amplitude. Exit-counting statistics therefore reveal how slow defects reorganize the timing of molecular production, a dynamical feature that mean current, mean density, and instantaneous polymerase occupancy cannot determine on their own.

\begin{acknowledgments}
No external funding was received for this work.
\end{acknowledgments}

\appendix

\section{Numerical methods and statistical analysis}
\label{app:numerical_methods}

\subsection{Continuous-time simulations}
\label{app:kinetic_monte_carlo}

An event-driven kinetic Monte Carlo algorithm simulates the open pausing TASEP exactly in continuous time. At a given configuration, the event list contains every allowed entrance, exit, forward hop, pausing transition, and unpausing transition. Their rates sum to the total escape rate
\begin{equation}
    R=R_{\rm in}+R_{\rm out}+R_{\rm hop}+R_{\rm pause}+R_{\rm unpause}.
\end{equation}
The time increment is drawn from an exponential distribution,
\begin{equation}
    \Delta t=-\frac{\ln u}{R},
\end{equation}
where $u$ is uniform on $(0,1)$. The algorithm then selects one allowed event with probability equal to its rate divided by $R$. This procedure includes rejected motion implicitly: a hop contributes to the event list only when its target site is empty.

Unless otherwise stated, the simulations use
\begin{equation}
    \epsilon=1,\qquad
    \alpha=0.1,\qquad
    \beta=1,\qquad
    k_u=10^{-3}.
    \label{eq:app_reference_parameters}
\end{equation}
The pausing rate $k_p$ and lattice length $L$ vary across the finite-size scans. These boundary rates place the pause-free reference process in an initiation-limited, non-exit-limited regime. The simulations therefore isolate the finite-size crossover generated by slow reversible pauses rather than remapping the full open-boundary phase diagram.

Each trajectory starts from an empty lattice. The main scans discard the rate-adapted transient
\begin{equation}
    t_{\rm burn}=\frac{100}{k_p}
    \label{eq:app_burn}
\end{equation}
and accumulate stationary observables over
\begin{equation}
    t_{\rm sample}=\frac{1000}{k_p}.
    \label{eq:app_sample}
\end{equation}
This protocol extends both relaxation and sampling when pauses become rarer. Each parameter point of the Fano-factor and finite-size scans contains ten independent replicates.

Time-integrated observables use the exact residence time of each configuration. The stationary current, total particle density, active-particle density, and paused-particle density therefore follow from time-weighted averages over the complete sampling interval. In particular,
\begin{equation}
    N_p=L\rho_{\rm paused}=\langle N_{\rm paused}\rangle
\end{equation}
is measured from the interacting stationary process rather than inferred from the isolated-particle paused fraction. Exit times are recorded after burn-in, and successive differences provide the inter-exit intervals.

\subsection{Window counts and Fano-factor estimation}
\label{app:fano_estimation}

For each trajectory, the sampling interval is divided into nonoverlapping windows. The analysis uses
\begin{multline}
 n_w=100,\ 200,\ 500,\ 10^3,\ 2\times10^3,\ 5\times10^3,\\
 10^4,\ 2\times10^4,\ 5\times10^4,\ 10^5,\ 2\times10^5,\ 5\times10^5
 \label{eq:app_window_numbers}
\end{multline}
windows, corresponding to observation times
\begin{equation}
    T_w=\frac{t_{\rm sample}}{n_w}.
\end{equation}
For window $j$, the exit count is
\begin{equation}
    N_{w,j}=N(t_j+T_w)-N(t_j).
\end{equation}
The sample mean and unbiased sample variance of these counts give
\begin{equation}
    \overline{N_w}=\frac{1}{n_w}\sum_{j=1}^{n_w}N_{w,j},
\end{equation}
\begin{equation}
    s^2_{N_w}=\frac{1}{n_w-1}
    \sum_{j=1}^{n_w}\left(N_{w,j}-\overline{N_w}\right)^2,
\end{equation}
and the trajectory-level Fano factor
\begin{equation}
    F(T_w)=\frac{s^2_{N_w}}{\overline{N_w}}.
\end{equation}
The analysis also records the fraction of windows containing no exit, which provides a diagnostic of sparse sampling at long $T_w$ or strongly inhibited parameter values.

Equation~\eqref{eq:telegraph} is fitted by nonlinear
ordinary least squares with $F_0$, $F_\infty$, and
$\tau_c$ free and constrained to positive values. No
uncertainty weighting is used in the production fits.
For the finite-size scans, the reported parameters are
obtained by fitting the Fano curve averaged across the
independent replicates. The fit parametrizes the crossover
between the short-window baseline and the long-window
plateau and does not impose an exact microscopic two-state
dynamics. Replicate-to-replicate dispersions quantify the
numerical variability of the underlying counting statistics.
The extraction of $F_\infty$ was tested against the long-window sampling and fitting procedure over all 57 conditions of the finite-size scan. Removing the largest observation window changed the fitted plateau by a median of $0.31\%$, with a $90$th-percentile absolute deviation of $1.83\%$; removing the two largest windows gave $0.51\%$ and $3.46\%$, respectively. Direct estimates obtained by averaging the two or three largest-window Fano factors closely followed the fitted plateaus ($r=0.998$ in both cases), with median deviations of about $-1.2\%$. These alternative estimates preserved the nonmonotonic dependence on $N_p$ and placed the maximum in the same order-one paused-population regime within the resolution of the parameter scans.

The finite-size maximum is located from the resolved local scan of $F_\infty$ as a function of $k_p$ or measured $N_p$. Only lengths with parameter points on both sides of the maximum enter Table~\ref{tab:maxima}. The quoted $k_p^{\rm max}$ is the simulated rate at the largest fitted plateau; no interpolation between neighboring rates is used.

\subsection{Unpausing-rate scan}
\label{app:ku_scan}

The unpausing-rate scan separates pause abundance from pause lifetime. It fixes $L=100$, varies $k_u$ from $5\times10^{-4}$ to $5\times10^{-3}$, and adjusts $k_p$ to obtain comparable measured paused populations around the three targets
\begin{equation}
    N_p^{\rm target}=1.5,\ 2,\ 3.
  \end{equation}
Each replicate is fitted separately to
Eq.~\eqref{eq:telegraph} by unweighted nonlinear least
squares, and the fitted parameters are then summarized
across replicates. The power laws plotted in
Fig.~\ref{fig:ku_scan} are effective fits over the simulated
interval; they summarize the observed dependence and are
not interpreted as asymptotic exponents.

\subsection{Initiation-rate control}
\label{app:alpha_control}

A separate scan tests whether the fluctuation maximum depends qualitatively on the reference initiation rate. The control fixes
\begin{equation}
L=100,\qquad \alpha=0.03,\qquad \beta=\epsilon=1,\qquad k_u=10^{-3},
\end{equation}
and varies
\begin{align}
k_p={}&1.0\times10^{-5},\,3.0\times10^{-5},\,6.0\times10^{-5},
\nonumber\\
&1.0\times10^{-4},\,1.6\times10^{-4},\,2.1\times10^{-4},
\nonumber\\
&3.3\times10^{-4},\,6.0\times10^{-4},\,1.1\times10^{-3}.
\end{align}
The simulation, relaxation, sampling, window-counting, and fitting procedures are otherwise identical to the main finite-size scan. Because congestion feeds back strongly on the total occupancy at low initiation, the measured $N_p$ rather than the nominal target labels defines the horizontal coordinate.

The scan spans measured paused populations from $N_p=0.038$ to $14.00$. The fitted plateau increases from $F_\infty=1.80$ at $N_p=0.038$ to a sampled maximum $F_\infty=47.25$ at $N_p=3.26$, then decreases to $4.69$ at $N_p=14.00$. The two neighboring conditions give $F_\infty=40.41$ at $N_p=1.57$ and $42.15$ at $N_p=4.51$, so the maximum is bracketed without requiring interpolation. The absolute fluctuation rate $JF_\infty$ is also nonmonotonic: it reaches its sampled maximum $JF_\infty=0.752$ at $N_p=1.57$ and decreases on either side. The lower initiation rate therefore changes both the amplitude and precise paused-population location of the relative-noise maximum but preserves the underlying intermediate, order-one fluctuation regime.

The total density changes substantially across this control, from $\rho=0.038$ in the nearly pause-free regime to approximately $0.28$ under strong congestion. This occupancy feedback explains why reducing $\alpha$ cannot be represented by a simple uniform rescaling of $k_p$: persistent pauses increase particle residence times and thereby raise the number of particles exposed to further pausing.
\subsection{Event-resolved output trajectories}
\label{app:output_traces}

The trajectories displayed in Fig.~\ref{fig:timeseries} use a separate protocol designed to resolve the temporal organization of exits. Five independent simulations are generated for each of three representative pausing rates. A fixed relaxation interval
\begin{equation}
    T_{\rm relax}=2\times10^4
\end{equation}
is followed by
\begin{equation}
    T_{\rm sample}=10^5=100k_u^{-1}
\end{equation}
of exit recording. Exact exit times are retained. The displayed current is binned over intervals of width
\begin{equation}
    \Delta=25,
\end{equation}
according to
\begin{equation}
    j_\Delta(t)=\frac{N(t+\Delta)-N(t)}{\Delta}.
\end{equation}
This bin width remains much shorter than the unpausing time $k_u^{-1}$ while suppressing event-by-event discreteness. The moving average over 50 bins serves only as a visual guide and does not enter any quantitative analysis.

\subsection{Structural sampling and cluster observables}
\label{app:structural_sampling}

The structural analysis uses $L=100$, the reference rates
in Eq.~\eqref{eq:app_reference_parameters}, ten independent
replicates, and the same burn-in and sampling durations as
the main Fano-factor scan. Structural summary statistics
are retained for all ten replicates. To limit output size,
the complete regularly sampled structural trajectory is
stored for replicate 1 at each parameter point. A targeted
reproducibility control stores four additional complete
trajectories at three representative conditions spanning
the fluctuation maximum, giving five independent detailed
trajectories at measured paused populations
$N_p\simeq0.73$, $1.95$, and $3.12$. Each detailed trajectory
is sampled at the regular interval
\begin{equation}
    \Delta t_{\rm struct}=100.
    \label{eq:app_struct_interval}
\end{equation}
The sampling interval is shorter than the characteristic
correlation time near the noise maximum, which is of order
$k_u^{-1}=10^3$, but it limits the resolution of residence
episodes shorter than $\Delta t_{\rm struct}$.

A particle cluster is a maximal sequence of consecutively occupied sites, irrespective of the internal state of its particles. For each sampled configuration, the analysis records the largest cluster size $C_{\max}$ and its position. It also records the first paused particle, the downstream-most paused particle, the cluster containing the downstream-most pause, and the upstream queue attached to that pause. The main structural comparison uses the threshold-free quantities
\begin{equation}
    \langle C_{\max}\rangle,
    \qquad
    {\rm Var}(C_{\max}).
\end{equation}
The binary indicator $C_{\max}>C_{\rm th}$ is used only for state-occupancy and residence-time diagnostics.

The Pearson coefficients quoted in the main text are descriptive correlations across the eight sampled parameter sets. They measure how closely two observables covary across this scan; they do not constitute an independent causal test.

\subsection{Pause-number statistics}
\label{app:pause_number}

The stored structural trajectory for replicate 1 at each
parameter point provides the instantaneous pause number
\begin{equation}
    M(t)=N_{\rm paused}(t).
\end{equation}
The empirical probabilities of the three classes used in Fig.~\ref{fig:pause_classes} are the fractions of sampled configurations satisfying
\begin{equation}
    M=0,\qquad M=1,\qquad M\geq2.
\end{equation}
Their Poisson reference values use the measured trajectory-averaged mean $N_p=\langle M\rangle$,
\begin{equation}
    P_{\rm Pois}(M=0)=e^{-N_p},
\end{equation}
\begin{equation}
    P_{\rm Pois}(M=1)=N_p e^{-N_p},
\end{equation}
\begin{equation}
    P_{\rm Pois}(M\geq2)=1-e^{-N_p}(1+N_p).
\end{equation}
The dispersion index
\begin{equation}
    D_M=\frac{{\rm Var}(M)}{\langle M\rangle}
\end{equation}
quantifies departures from the Poisson variance-to-mean relation.

Temporal correlations between successive structural samples
make independent-sample binomial errors inappropriate.
Uncertainties in the pause-number statistics are therefore
estimated from the stored detailed trajectory using a
circular moving-block bootstrap with 300 resamples and
random seed 20260729. For each parameter point, the block
length is set to twice the estimated integrated
autocorrelation time of $N_{\rm paused}$, expressed in
units of the structural-sampling interval. Each bootstrap
realization resamples contiguous blocks, thereby preserving
short-range temporal dependence while recomputing the
pause-class probabilities and related statistics. The
resulting bootstrap dispersion provides the error bars
reported in Fig.~\ref{fig:pause_classes}.

The independent five-trajectory control confirms that the
pause-number statistics are reproducible across trajectories.
At the condition closest to the output-noise maximum,
$N_p=1.946\pm0.024$ (mean $\pm$ standard deviation across
trajectories), the replicate-averaged probabilities are
$P(M=0)=0.291\pm0.004$, $P(M=1)=0.234\pm0.004$, and
$P(M\geq2)=0.474\pm0.006$. The dispersion index remains
larger than unity in all five trajectories and increases
across the three control conditions, with replicate means
$D_M=1.74\pm0.02$, $2.08\pm0.03$, and $2.34\pm0.03$ for
$N_p\simeq0.73$, $1.95$, and $3.12$, respectively. These
between-trajectory controls confirm that the deviations from
the Poisson reference do not originate from the single
trajectory used to draw Fig.~\ref{fig:pause_classes}.

\subsection{Residence-time and telegraph analyses}
\label{app:residence_times}

Residence episodes are reconstructed from the stored
regularly sampled structural trajectory for replicate 1
at each parameter point. Two alternative binary partitions are considered. The pause-number partition distinguishes
\begin{equation}
    U_0:\ M=0,
    \qquad
    P:\ M\geq1,
\end{equation}
while the congestion partition distinguishes
\begin{equation}
    U:\ C_{\max}\leq C_{\rm th},
    \qquad
    B:\ C_{\max}>C_{\rm th}.
\end{equation}
A residence episode consists of consecutive samples assigned to the same state. Its duration equals the number of consecutive sampling intervals multiplied by $\Delta t_{\rm struct}$. Episodes truncated by the beginning or end of the sampled trajectory are excluded from the residence-time averages and survival curves, so that incomplete durations do not bias the estimated means.

For a binary continuous-time Markov process with mean residence times $\tau_1$ and $\tau_2$, the transition rates are $1/\tau_1$ and $1/\tau_2$, and the correlation time is
\begin{equation}
    \tau_{\rm tel}
    =\frac{1}{\tau_1^{-1}+\tau_2^{-1}}
    =\frac{\tau_1\tau_2}{\tau_1+\tau_2}.
    \label{eq:app_telegraph_time}
\end{equation}
The pause-number estimate uses $\tau_1=\tau_0$ and $\tau_2=\tau_P$ and is denoted $\tau_{\rm pause}$. The congestion estimate uses $\tau_1=\tau_U$ and $\tau_2=\tau_B$ and is denoted $\tau_{\rm tel}$.

Uncertainties in the residence-time estimates are obtained
from 300 moving-block bootstrap resamples of the ordered
residence-episode sequence, using random seed 20260729.
The block length is chosen separately for each condition
so that a block spans approximately two integrated
autocorrelation times; the mean episode duration converts
this correlation time into a number of successive episodes.

The threshold analysis spans
\begin{equation}
    \frac{C_{\rm th}}{L}=0.15,\ 0.20,\ 0.25,\ 0.30.
\end{equation}
Figure~\ref{fig:telegraph_residence} uses $C_{\rm th}/L=0.20$ as a representative value and displays the full threshold sensitivity through the ratios $\tau_{\rm tel}/\tau_c$. The survival function of a residence time $T$ is estimated empirically as
\begin{equation}
    S(t)=P(T>t).
\end{equation}
The exponential references in Fig.~\ref{fig:telegraph_residence} use the measured mean residence times,
\begin{equation}
    S_{\rm exp}(t)=e^{-t/\tau}.
\end{equation}
These curves test whether a Markov telegraph process provides a useful effective description. Deviations from exponential survival and sensitivity to the state definition prevent an interpretation as an exact microscopic reduction.

The targeted five-trajectory control tests the central
timescale comparison independently of the trajectory used in
Fig.~\ref{fig:telegraph_residence}. At
$N_p\simeq0.73$, $1.95$, and $3.12$, the mean replicate-wise
ratios are
$\tau_{\rm pause}/\tau_c=0.83\pm0.27$,
$0.79\pm0.19$, and $0.80\pm0.17$, respectively, where the
uncertainties are standard deviations across five independent
trajectories. The pause-based ratio therefore remains on the
same scale without systematic drift across the crossover. By
contrast, for the representative threshold
$C_{\rm th}/L=0.20$, the corresponding ratios
$\tau_{\rm tel}/\tau_c=0.96\pm0.31$,
$1.24\pm0.31$, and $1.83\pm0.37$ increase with $N_p$.
Independent replication thus preserves the separation seen
in Fig.~\ref{fig:telegraph_residence}: pause-number dynamics
tracks the output-decorrelation scale, whereas a fixed
cluster threshold progressively separates from it as
congestion becomes persistent.

\subsection{Uncertainty and reproducibility}
\label{app:uncertainty}

Independent simulation replicates define the primary unit
of numerical replication for the finite-size, initiation-rate,
unpausing, and structural summary observables. Quantities derived
directly from complete replicate outputs are first computed
separately and then summarized across replicates. The full
structural scan uses one stored detailed trajectory per
parameter point and moving-block-bootstrap uncertainties that
preserve temporal dependence. The targeted control at
$N_p\simeq0.73$, $1.95$, and $3.12$ adds four independently
generated detailed trajectories to the original trajectory
at each condition. Standard deviations across these five
trajectories quantify between-trajectory reproducibility of
the pause-number probabilities and residence-time ratios.

The reproducibility package prepared for public release contains the simulation parameters,
replicate and aggregate Fano summaries, stationary
observables, structural replicate summaries, the stored
structural trajectories used for
Figs.~\ref{fig:pause_classes} and
\ref{fig:telegraph_residence}, the additional independent
structural trajectories used for the reproducibility controls,
and the event-resolved exit data used for
Fig.~\ref{fig:timeseries}, together with the analysis and
figure-generation scripts.

\end{document}